\documentclass[prl,aps,twocolumn,amsfonts,showpacs,superscriptaddress,floatfix]{revtex4} 
\usepackage{epsfig}
\usepackage{psfrag}
\usepackage{color}
\usepackage{graphicx}
\usepackage{subfigure}

\def\be{\begin{equation}}
\def\ee{\end{equation}}
\def\bea{\begin{eqnarray}}
\def\eea{\end{eqnarray}}

\begin{document}
\title{Itineracy effects on spin correlations in 1D Mott insulators}
\author{M. J. Bhaseen} 
\affiliation{Department of Physics, Brookhaven National Laboratory,
  Upton, NY 11973-5000, USA} 
\author{F. H. L. Essler}
\affiliation{Department of Physics, Brookhaven National Laboratory,
  Upton, NY 11973-5000, USA} 
\author{A. Grage}
\affiliation{Fachbereich Physik, Philipps-Universit\"at Marburg, D-35032 Marburg, Germany}
\date{\today}

\begin{abstract}
We consider spin correlations in the one dimensional half filled
repulsive Hubbard model. For very large values of the on-site repulsion
$U$ the spin correlations are dominated by virtual hopping processes
of electrons and are described in terms of a spin-1/2 Heisenberg
chain. As $U$ is decreased real hopping processes of electrons become
important and eventually dominate the spin response. We discuss the
evolution of the dynamical structure factor as a function of $U$. We comment
on the relevance of our results for inelastic neutron scattering
experiments on quasi-1D materials.
\end{abstract}

\pacs{71.10.Pm, 71.10.Fd, 72.80.Sk}

\maketitle

Quasi-one-dimensional (Q1D) Mott insulators are a paradigm for the
importance of strong correlations and are known to exhibit a wide
variety of unusual physical phenomena such as spin-charge separation
and quantum number fractionalization. Experimental realizations include
the chain cuprates ${\rm SrCuO}_2$ and ${\rm Sr}_2{\rm Cu}{\rm  O}_3$
and a number of organic compounds \cite{Ishiguro:Organic,Bourbannis:Normal}. It has been known for a long time that the magnetic properties of Q1D
Mott insulators are rather unusual. The dynamical structure factor is
entirely incoherent and reflects the fact that the elementary
spin excitations carry spin $1/2$. They are very different in nature
from antiferromagnetic spin waves.
In the limit of very strong Coulomb repulsion the spin degrees of
freedom are commonly modelled by the spin-1/2 Heisenberg chain. This 
has proved to provide an adequate description of recent inelastic
neutron scattering experiments on ${\rm SrCuO_2}$ \cite{zaliznyak99,zaliznyak03}.
In this limit the low energy dynamics is dominated by spin flips
generated through virtual hopping processes of electrons. However, for general
Mott insulating materials there is no reason for the Coulomb repulsion
to be much stronger than the electron hopping amplitude. Indeed, this situation is realized in the Bechgaard salts. A natural
question to ask then is how electron itineracy affects the spin dynamics. This issue has recently attracted much attention in the
context of the  two-dimensional Mott insulating cuprate ${\rm
  La}_2{\rm   Cu}{\rm   O}_4$ \cite{Coldea:Spinwave}. The
experimentally observed spin wave dispersion was found to depart
significantly from the Heisenberg form, revealing the presence of ring-exchange interactions. This is a direct manifestation of electron
itineracy in the sense that ring-exchange interactions are generated
by higher order virtual electron hopping processes.

Motivated by these findings we investigate the effects of
electron itineracy on the unusual spin correlations in Q1D Mott
insulators. We do so in the simplest model that incorporates the
relevant physics, the one dimensional half filled Hubbard model.
The Hamiltonian is given by

\begin{eqnarray}
H &=&-t\sum_{l,\,\sigma}(c_{l,\sigma}^{+}c_{l+1,\sigma}+c_{l+1,\sigma}^{+}c_{l,\sigma}) \nonumber \\
  & &+U\sum_{l}(n_{l,\uparrow}-\frac{1}{2})(n_{l,\downarrow}-\frac{1}{2}), \label{fermihamil}
\end{eqnarray}
where $c^{+}_{l,\sigma}$  creates an electron with spin $\sigma=\uparrow,\downarrow$ at site $l$,  $n_{l,\sigma}\equiv c^+_{l,\sigma}c_{l,\sigma}$, and $U$ is the on-site Coulomb repulsion. In momentum space, the kinetic term  yields a cosine band, $\varepsilon(k)=-2t\cos(ka_0)$, where $a_0$ is the lattice spacing. At half-filling, $k_F=\pm\pi/2a_0$. The elementary excitations of this model are termed spinons and holons. The spinons are gapless, uncharged, and carry spin $1/2$. Their energy and momentum are conveniently parameterized in terms of the spectral parameter $\Lambda$ \cite{Ovchinnikov:Excitation}: 
\bea
E_s(\Lambda) & = & 2t\int_{0}^{\infty}\frac{dx}{x}\frac{J_1(x)\cos(x\Lambda)}{\cosh(x U/4t)}, \label{espinon} \\
P_s(\Lambda) &= & \frac{1}{a_0}\left[\frac{\pi}{2}-\int_0^\infty \frac{dx}{x}\frac{J_0(x)\sin(x\Lambda)}{\cosh(xU/4t)}\right], \label{pspinon}
\eea
where $J_0(x)$ and $J_1(x)$ are Bessel functions, and $E_s(0)$ plays the role of a ``spinon bandwidth''. The holon and antiholon, are gapped, spinless, and carry charge $|e|$ and $-|e|$ respectively. The ``charge gap'' $\Delta$, defined as the minimum of the holon energy, increases with $U$ and is given by:
\be
\Delta=-2t+\frac{U}{2}+2t\int_0^\infty \frac{dx}{x}\frac{J_1(x)e^{-xU/4t}}{\cosh(xU/4t)}.
\ee
 These excitations form a basis of scattering states \cite{Essler:both} in which one may explore the dynamical response of 1D Mott insulators.

\begin{figure}
\begin{center}
\subfigure[$U\gg t$]
{
\psfrag{q}{$q$}
\psfrag{e}{$\rotatebox{180}{$\omega/J$}$}
\psfrag{1}{$\frac{\pi}{2}$}
\psfrag{2}{$\pi$}
\psfrag{0}{$0\,\,$}
\psfrag{Pi}{$\,\pi$}
\psfrag{2 Pi}{$\quad2\pi$}
\label{fig:a}
\includegraphics[height=3.7cm]{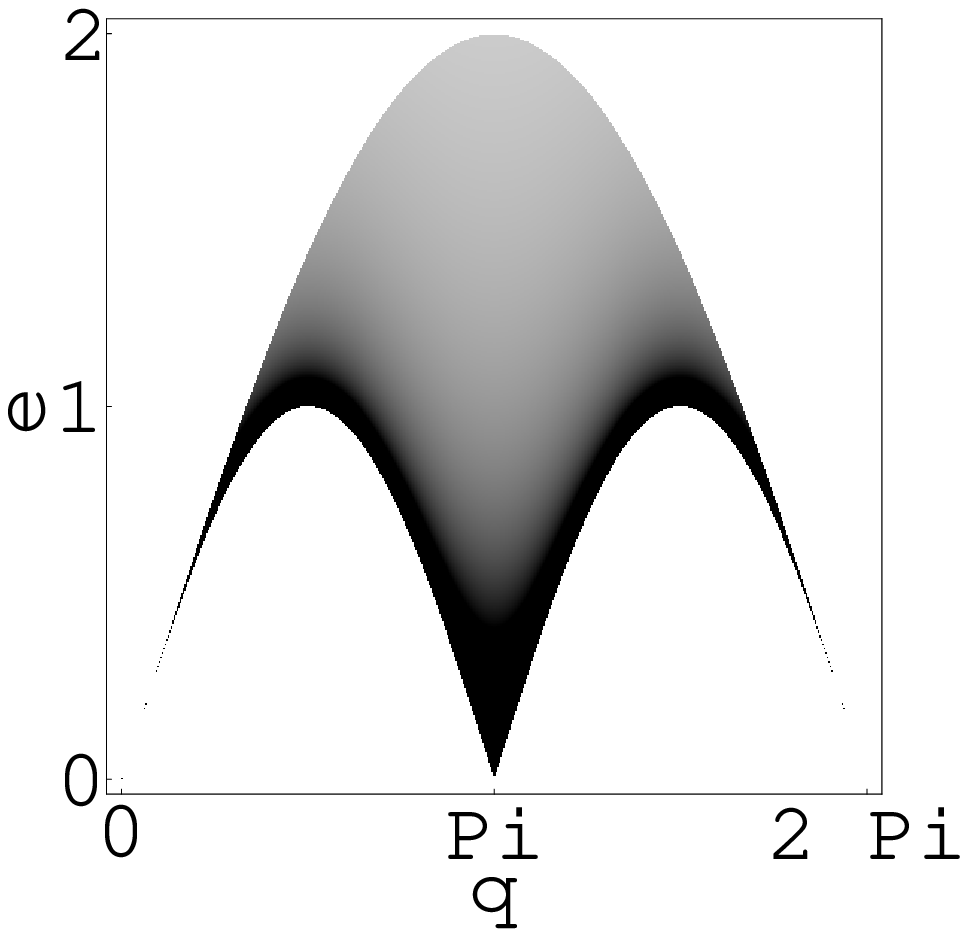}
}
\subfigure[$U\ll t$]
{
\psfrag{q}{$q$}
\psfrag{e}{$\rotatebox{180}{$\omega/t$}$}
\psfrag{2}{$2$}
\psfrag{4}{$4$}
\psfrag{0}{$0\,\,$}
\psfrag{Pi}{$\,\pi$}
\psfrag{2 Pi}{$\quad2\pi$}
\label{fig:b}
\includegraphics[height=3.7cm]{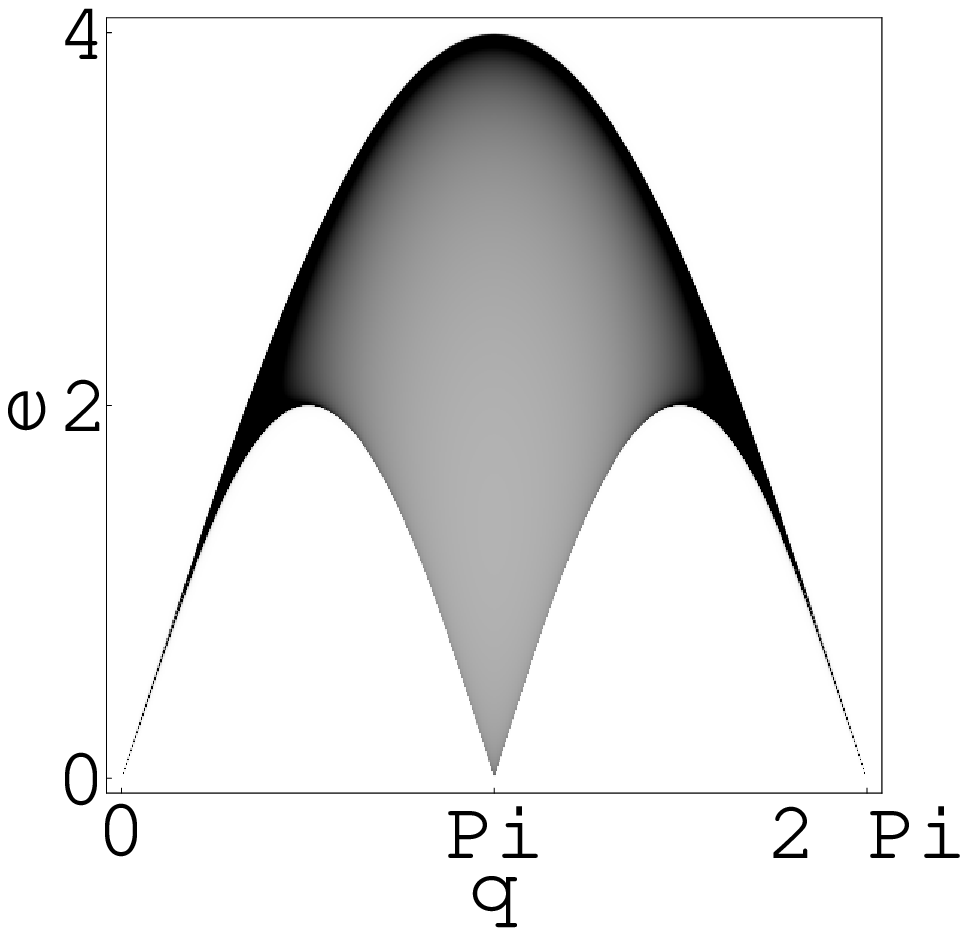}
}
\end{center}
\caption{\label{fig:} Density plots of the dynamical structure factor: $(a)$ $U\gg t$ M\"uller approximation, $(b)$ $U\ll t$ result for $U/t=0.4$. At large (small) $U/t$ the spectral weight is on the lower (upper) spinon boundary.}
\end{figure}

 As usual, the dynamical structure factor is obtained from the imaginary part of the Fourier transform of the spin-spin correlation function. In view of the ${\rm SU}(2)$ spin symmetry of the Hamiltonian we consider:
\be
S^{zz}(\omega,q)=-{\rm Im}\,\left(\chi_E^{zz}(\bar\omega,q)|_{\bar\omega\rightarrow\eta-i\omega}\right),
\ee
where $\eta\rightarrow 0^+$ and the susceptibility (in Euclidean time) is given by
\bea
\chi_E^{zz}(\bar\omega,q)= -\int_{-\infty}^{\infty} dx\ d\tau 
\,e^{i\bar\omega\tau-iqx} \langle\, S^z(\tau,x)S^z(0,0)\,\rangle.
\eea
The spin operator is given by $S^z\equiv (n_{\uparrow}-n_{\downarrow})/2$. In the first instance, it is instructive to study the known limits of large and small $U/t$. In the limit $U \gg t$, the half filled Hubbard model reduces to an isotropic Heisenberg antiferromagnet with exchange $J\simeq 4t^2/U$ \cite{Anderson:J}. In this case, the structure factor is well approximated by the so-called M\"uller ansatz \cite{Muller:Ansatz}:
\be
S^{zz}(\omega,q)\propto\frac{\Theta(\omega-\omega_L)\Theta(\omega_U-\omega)}{\sqrt{\omega^2-\omega_L^2}}; \quad U\gg t.
\ee
Whilst more detailed results can been obtained using integrability \cite{Jimbo:Algebraic} this approximation is sufficient for our current discussion. The lower and upper spinon boundaries, $\omega_L=(\pi/2)J\,|\sin qa_0|$ and $\omega_U=\pi J\,|\sin(qa_0/2)|$, follow from the spinon dispersion relations (\ref{espinon}) and (\ref{pspinon}). These boundaries delimit a continuum of two spinon excitations, and we plot this in Fig. \ref{fig:a}. It is readily seen that the spectral weight (plotted darker) is concentrated on the {\it lower} spinon boundary. The power law singularity extends down to $\omega=0$ in the vicinity of $q=\pi$, and reflects the gapless nature of spinons. 

\begin{figure}
\begin{center}
\noindent
\includegraphics[angle=270,width=8cm]{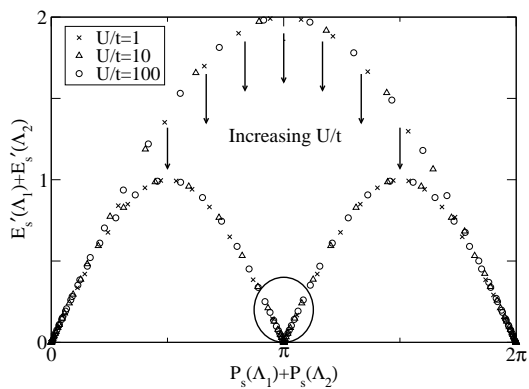}
\end{center}
\caption{\label{fig:scaled} Approximate collapse of the rescaled
  spinon continuum.  We plot the rescaled spinon energy
  $E_s^\prime(\Lambda)\equiv E_s(\Lambda)/E_s(0)$ versus momentum
  $P_s(\Lambda)$. The arrows show the transfer of spectral
  weight as $U/t$ is increased.}   
\end{figure}

In the limit of large external freqencies and $U \ll t$, one may treat the Hubbard interaction perturbatively. Using the tightbinding propagator ${\mathcal G}(\bar\omega,q)=a_0[i\bar\omega+2t\cos q]^{-1}$, a summation over ``bubble'' diagrams yields:
\be
\label{bubblesum}
\chi_E^{zz}(\bar\omega,q)=\frac{\chi_{0\,E}^{zz}(\bar\omega,q)}{1+2(U/a_0)\chi_{0\,E}^{zz}(\bar\omega,q)}; \quad U\ll t,
\ee
where the $U=0$ single bubble is given by
\be
\chi_{0\,E}^{zz}(\bar\omega,q)=-\frac{a_0}{\pi}\frac{{\rm arth}\,\left[\frac{4t(\sin qa_0/2)^2}{\sqrt{\bar\omega^2+(4t\sin qa_0/2)^2}}\right]}{\sqrt{\bar\omega^2+(4t\sin qa_0/2)^2}}.
\ee
We plot the corresponding structure factor in Fig. \ref{fig:b}. In this limit, significant spectral weight is concentrated on the {\it upper} spinon boundary. (It is evident that this large $\omega$ approximation fails to capture the low frequency spinon divergence; see discussion below.)
Thus apart from a rescaling of the spinon bandwidth, to which we shall return below and in Fig. \ref{fig:scaled}, there is a manifest {\it redistribution} of spectral weight as one changes the ratio $U/t$. It is therefore highly desirable to build up a better picture for what happens in the intermediate regime $U\sim t$. We address this pertinent issue using the techniques of integrable field theory.

At low energies, we may linearize the non-interacting fermionic spectrum around $\pm |k_F|$.  The resulting model may be bosonized \cite{Tsvelik:boson} in terms of two Bose fields, $\Phi_s$ and $\Phi_c$. These correspond to the collective spin and charge degrees of freedom respectively. 
The continuum Hamiltonian density separates in two commuting pieces: 
\bea
{\mathcal H}  & =  &{\mathcal H}_s+{\mathcal H}_c, \nonumber
\eea
where
\bea
\mathcal{H}_{s} & = &  \frac{v_s}{16\pi} \left[(\partial_x\Phi_s)^2-v_s^{-2}(\partial_\tau\Phi_s)^2\right]-g\,:{\bf J}\cdot {\bar {\bf J}}:,\nonumber \\
\mathcal{H}_{c} & = &  \frac{v_c}{16\pi} \left[(\partial_x\Phi_c)^2-v_c^{-2}(\partial_\tau\Phi_c)^2\right]+g\,:{\bf I}\cdot {\bar {\bf I}}:. \nonumber 
\eea
Here ${\bf J}$ and ${\bf I}$ are the ${\rm SU}(2)$ spin and isospin currents, $v_s=v_F-Ua_0/2\pi$, $v_c=v_F+Ua_0/2\pi$ and $g=2Ua_0$. The interaction in the charge sector is marginally relevant and leads to the formation of the charge gap. The interaction in the spin sector is marginally irrelevant and will be neglected. This description is valid for $\Delta\ll 4t$ and for $U\alt 3t$.
In order to enhance the clarity of presentation, we will mainly work in the so-called scaling limit:
\bea
t\rightarrow\infty,\quad U/t\rightarrow 0,\quad \Delta\rightarrow m\equiv \frac{4}{\pi}\sqrt{Ut}\,e^{-2\pi t/U}\,\,  {\rm fixed}. \nonumber
\eea
In this limit $v_s$ coincides with $v_c$ and Lorentz invariance emerges. The charge sector is governed by the ${\rm SU}(2)$ invariant sine Gordon model, and the spin sector by a massless Gaussian model. This simplifies the ensuing formulae, whilst capturing the relevant physics. Where appropriate, we give the results for $v_s\neq v_c$.   

\begin{figure}
\includegraphics[angle=270,width=8.5cm]{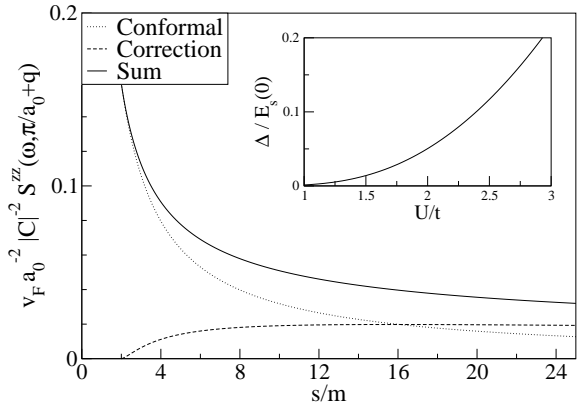}
\caption{\label{fig:itincorr}The staggered dynamical structure
  factor. We plot the conformal contribution, the holon-antiholon
  ``itineracy correction'' and their sum. The inset shows the ratio of
  the charge gap to the spinon bandwidth as a function of $U/t$.} 
\end{figure}

In this low energy description, the $z$ component of the spin operator takes the form:
\be
\label{zcomp}
S^z=-\frac{a_0}{4\pi}\partial_x\Phi_s+(-1)^{\frac{x}{a_0}}\frac{a_0}{\pi}\,\cos\frac{\Phi_c}{2}\,\sin\frac{\Phi_s}{2}.
\ee
We draw attention to the fact that both the ``spin'' {\it and} the
``charge'' fields enter the staggered component of the spin
operator. Excitations of the charge field, and thus electron
itineracy, may therefore influence magnetic correlations in the
vicinity of $q\sim\pi$. In the first, crude approximation, the
operator $\cos\frac{\Phi_c}{2}$ is replaced by its constant vacuum
expectation value $\langle \,0\,|\cos\frac{\Phi_c}{2}\,|\,0\,\rangle$,
and correlators are given by the critical spin sector alone. However,
this is merely the leading term in a systematic expansion of the
charge sector. In contrast to naive expectations, these terms are
shown to be significant. In the subsequent analysis,  we shall
calculate the ``itineracy corrections'' to the magnetic structure
factor arising from these charge excitations. We employ a number of
results developed in the study of density-density correlations
\cite{Density:Controzzi}. As follows from equation (\ref{zcomp}) the
staggered part of the spin-spin correlation function is given by 
\be
\langle \,S^z(\tau,x)S^z(0,0)\,\rangle_{{\rm stagg.}}=(-1)^{\frac{x}{a_0}}\frac{1}{2\pi^2}\frac{a_0^2}{v_F}\frac{{\mathcal F}(r)}{r},
\ee
where $r=\sqrt{\tau^2+x^2/v_F^2}$ and we have defined the contribution from the gapped charge sector
\be
\label{chargecorr}
{{\mathcal F}}(r)\equiv \langle\, 0\,|\cos\frac{\Phi_c}{2}(\tau,x)\cos\frac{\Phi_c}{2}(0,0)|\,0\,\rangle.
\ee
We introduce a basis of states $|\,\theta_n\dots\theta_1\,\rangle_{\epsilon_n\dots\epsilon_1}$ where the index $\epsilon=\pm$ denotes holon and antiholon, and $\theta$ are their rapidities. Inserting the resolution of the identity
\bea
& & 1  = |\,0\,\rangle\langle\,0\,| + \nonumber \\
&  & \sum_{n=1}^\infty\sum_{\epsilon_i}\int\frac{d\theta_1\dots d\theta_n}{(2\pi)^n n!}\,|\,\theta_n\dots\theta_1\,\rangle_{\epsilon_n\dots\epsilon_1} \ ^{\epsilon_n\dots\epsilon_1}\langle\,\theta_n\dots\theta_1\,|\nonumber 
\eea
one may develop (\ref{chargecorr}) in a systematic expansion over multi-particle intermediate states:
\be
\label{realspaceexpand}
{{\mathcal F}}(r)={{\mathcal F}}_0(r)+{{\mathcal F}}_2(r)+\dots
\ee
The leading term is obtained from the vacuum expectation value of the operator
\be
{{\mathcal F}}_0(r)=\left|\langle\,0\,| \cos\frac{\Phi_c}{2}|\,0\,\rangle\right|^2\equiv \frac{m}{v_F}\,|{\mathcal C}|^2
\ee
and is a constant. The dimension of $\langle\,0\,| \cos\frac{\Phi_c}{2}|\,0\,\rangle$ is $[L]^{-1/2}$. The dimensionless constant ${\mathcal C}$ is pinned by the short distance properties of the charge sector \cite{Lukyanov:Expectation} and fixes the normalization of the structure factor. That is to say, to leading order, the staggered spin-spin correlation function is approximated by the conformal $1/r$ decay.  Since the operator $\cos\Phi_c/2$ does not couple to single holon states the next contribution may be written:
\bea
{{\mathcal F}}_2(r) & = & \sum_{\epsilon_1,\epsilon_2}\int\frac{d\theta_1\, d\theta_2}{(2\pi)^2 2!}\,\left|\,\langle\,0\,|\cos\frac{\Phi_c}{2}\,|\,\theta_2\,\theta_1\,\rangle_{\epsilon_2,\epsilon_1}\,\right|^2  \nonumber \\
& & \hspace{-1.7cm}\times \exp\left[-m({\rm ch}\,\theta_1+{\rm ch}\,\theta_2)|\tau|-im({\rm sh}\,\theta_1+{\rm sh}\,\theta_2)x/v_F\right].  \nonumber 
\eea
Defining $\theta=\theta_1-\theta_2$, the form factor is given by
\cite{Smirnov:Form,Lukyanov:Exponential} 
\bea
f^{\,\cos\frac{\Phi_c}{2}}(\theta)_{+-} &\equiv&
\langle\,0\,|\cos\frac{\Phi_c}{2}\,|\,\theta_2\,\theta_1\,\rangle_{+-}
\nonumber\\
&=&\sqrt{{\mathcal F}_0}
\,\,\frac{2i\ {\rm sh}(\theta/2)}{\theta+i\pi}\ E(\theta)\ ,
\eea
\be
E(\theta)=\exp\left(-\int_0^\infty\frac{dx}{x}
\frac{\sin^2([\theta+i\pi]x/\pi)e^{-x}}{{\rm sh}(2x){\rm ch}(x)}\right).
\ee
In view of (\ref{realspaceexpand}) we may expand the structure factor
\be
S^{zz}(\omega,q)   =\sum_{n=0}^\infty S_{2n}^{zz}(\omega,q)\ ,
\ee
where $S_{2n}^{zz}(\omega,q)$ denotes the contribution due to
intermediate states with $2n$ (anti)holons. The ``conformal''
contribution is easily evaluated:
\be
S_0^{zz}(\omega,\frac{\pi}{a_0}+q) = \frac{a_0^2}{v_F}\frac{|\,{\mathcal
    C}|^2\,}{\pi}\frac{\Theta(y)}{y};\quad y\equiv\frac{s}{m}, 
\ee
where $\Theta(y)=1$ for $y\ge 0$ is the Heaviside step function, and
$s\equiv\sqrt{\omega^2-v_F^2\,q^2}$. Performing the sum over isotopic
indices and the necessary integrals \cite{Density:Controzzi}, the
``itineracy correction'' is found to be 
\bea
\label{itincorr}
&&{S}_2^{zz}(\omega,\frac{\pi}{a_0}+q)   =\frac{a_0^2}{v_F}
\frac{|\,{\mathcal C}|^2}{2\pi^2}\nonumber\\
&&\times\ \int_{0}^{\theta^\prime}
\frac{d\theta}{\gamma}\,\frac{8\,{\rm sh}^2\theta}{4\theta^2+\pi^2}\,
|E(\theta)|^2\,F\left[\frac{1}{2},\frac{1}{2};1;1-\frac{y^2}{\gamma^2}
\right],\quad
\eea
where $\theta^\prime=\mathrm{arch}\,(s/2m)$, $\gamma\equiv2\,{\rm ch}\,\theta$.
We plot these contributions to the structure factor in
Fig. \ref{fig:itincorr}.
For $v_s\neq v_c$ (\ref{itincorr}) may be generalized to
\bea
{S}_2^{zz}(\omega,\frac{\pi}{a_0}+q)  & =  & \nonumber \\
& & \hspace{-3cm} \frac{a_0^2}{2\pi^2}\int_{-\infty}^{\infty}
\frac{d\theta  \,d\theta^\prime}{\pi}\,
\frac{|f^{\,\cos\frac{\Phi_c}{2}}(2\theta)_{+-}|^2}
{\sqrt{\Omega_+\,\Omega_-}}\,\Theta(\Omega_+)\,\Theta(\Omega_-),
\nonumber   
\eea
where $\Omega_\pm=\omega\pm v_sq-2\Delta\,{\rm ch}\,\theta\,({\rm
  ch}\,\theta^\prime\pm (v_s/v_c)\,{\rm sh}\,\theta^\prime)$.
In the vicinity of the threshold ($s=2m$), the integral in
(\ref{itincorr}) may be obtained by
Taylor expanding the integrand. The leading threshold behavior as
$s\rightarrow 2m^{+}$ is 
\be
S_2^{zz}(\omega,\frac{\pi}{a_0}+q)\rightarrow
\frac{a_0^2}{v_F}\frac{|\,{\mathcal
    C}|^2}{2\pi^2}\frac{4|E(0)|^2}{3\pi^2}\,\left[\frac{s-2m}{m}\right]^{3/2}.  
\ee 
In contrast to the conformal contribution, this term goes to zero at
its threshold and is free of singularities. At high energies $s\gg
2m$, $S_2^{zz}(\omega,\frac{\pi}{a_0}+q)\rightarrow{\rm constant.}$ In real
space, the asymptotics of this correction (evaluated in the saddle
point approximation) yield: 
\bea
\langle \, S^z(\tau,x)S^z(0,0)\,\rangle_{\rm stagg.} & = & \nonumber \\
& & \hspace{-3.5cm} (-1)^{\frac{x}{a_0}}\frac{|\,{\mathcal
    C}|^2}{2\pi^2}\frac{a_0^2m}{v_F^2r}\left[1+{\mathcal A}\,\frac
  {e^{-2mr}}{(mr)^2}+\dots \right],  
\eea
for $(mr)\gg 1$, where
\be
{\mathcal   A}=\frac{2\sqrt{2}}{\pi^3}
\exp\left(-1+\int_0^\infty\frac{dx}{x}\frac{e^{-x}\,{\rm sh}\,x}{{\rm ch}^2\,x}\right). 
\ee

It is apparent from Fig. \ref{fig:itincorr}, that the itineracy
correction becomes significant for $s\sim 10m$; for $v_s\neq v_c$,
$s\sim 10\Delta$. In order to observe this feature within the two
spinon continuum, we seek to maximize $\Delta$ subject to  
$10 \Delta \alt 2 E_s(0) \ll 40 t$.
As may be seen from the inset of Fig. \ref{fig:itincorr}, this
condition is satisfied for $2\alt U/t\alt 3$. 
In particular, the spinon bandwidth $E_s(0)$ decreases with increasing
$U/t$. In view of this we rescale the spinon energies by this
factor. As shown in Fig. \ref{fig:scaled}, the rescaled dispersion
relations collapse (approximately, but remarkably well) onto a single
curve. We may therefore compare systems with different values of $U/t$
on a single plot. 
In conjunction with the limiting cases, $U\ll t$ and $U\gg t$, our
detailed analytic results support the simple yet compelling view
depicted in Fig. \ref{fig:scaled}. As $U/t$ increases spectral weight
is transferred from the upper rescaled boundary, and builds up in both
the conformal divergence and the itineracy correction. The downward
shift at high frequencies is evident from the bubble summation
(\ref{bubblesum}) and the build up at low frequencies originates in
the increase of the charge gap.  

In this letter we have determined dynamical spin correlations
in 1D Mott insulators. Away from the well-understood
``Heisenberg limit'' of very large Mott gaps, electron itineracy
effects are shown to be quite significant. On the basis of our
calculations we have put forward a simple picture describing the
spectral weight transfer in the dynamical structure factor as a
function of the strength of the Coulomb repulsion. We expect our
findings to be of relevance for inelastic neutron scattering
experiments on quasi-1D small-gap Mott insulators.

We thank F. Gebhard, A.M. Tsvelik and especially I. Zaliznyak for
important discussions. This work was supported by the U.S. DOE under 
Contract No DE-AC02-98 CH10886. AG acknowledges support by the Theory 
Institute for Strongly Correlated and Complex Systems at BNL and
the Optodynamics Center of the Universit\"at Marburg.

\bibliographystyle{h-physrev}

\end{document}